\documentclass[10pt]{llncs} 

\usepackage{graphicx}
\usepackage{etoolbox}
\usepackage{hyperref}
\usepackage{float}
\usepackage[utf8]{inputenc}
\usepackage{array,colortbl,xcolor}
\usepackage{listings}
\usepackage{bera}
\usepackage{xcolor}

\colorlet{punct}{red!60!black}
\definecolor{background}{HTML}{EEEEEE}
\definecolor{delim}{RGB}{20,105,176}
\colorlet{numb}{magenta!60!black}

\lstdefinelanguage{json}{
    basicstyle=\normalfont\ttfamily,
    numbers=left,
    numberstyle=\scriptsize,
    stepnumber=1,
    numbersep=8pt,
    showstringspaces=false,
    breaklines=true,
    frame=lines,
    backgroundcolor=\color{background},
    literate=
     *{0}{{{\color{numb}0}}}{1}
      {1}{{{\color{numb}1}}}{1}
      {2}{{{\color{numb}2}}}{1}
      {3}{{{\color{numb}3}}}{1}
      {4}{{{\color{numb}4}}}{1}
      {5}{{{\color{numb}5}}}{1}
      {6}{{{\color{numb}6}}}{1}
      {7}{{{\color{numb}7}}}{1}
      {8}{{{\color{numb}8}}}{1}
      {9}{{{\color{numb}9}}}{1}
      {:}{{{\color{punct}{:}}}}{1}
      {,}{{{\color{punct}{,}}}}{1}
      {\{}{{{\color{delim}{\{}}}}{1}
      {\}}{{{\color{delim}{\}}}}}{1}
      {[}{{{\color{delim}{[}}}}{1}
      {]}{{{\color{delim}{]}}}}{1},
}

\restylefloat{table}

\patchcmd{\quote}{\rightmargin}{\leftmargin 0.5em \rightmargin}{}{}
\graphicspath{ {images/} }

\title{An architecture for non-invasive \\ software measurement}

 \author{Vasilii Artemev\inst{2}, Vladimir Ivanov\inst{1}, Manuel Mazzara\inst{1}, Alan Rogers\inst{1} \\ Alberto Sillitti\inst{1}, Giancarlo Succi\inst{1} and Eugene Zouev\inst{1}}
 
\institute{
Innopolis University, Russian Federation \\
\email{\{v.ivanov, m.mazzara, a.rogers, g.succi, e.zuev\}@innopolis.ru}
\and 
\email{vasart@gmail.com}
}

\begin{document}

\maketitle

\newlength\Origarrayrulewidth
\newcommand{\Cline}[1]{%
  \noalign{\global\setlength\Origarrayrulewidth{\arrayrulewidth}}%
  \noalign{\global\setlength\arrayrulewidth{2pt}}\cline{#1}%
  \noalign{\global\setlength\arrayrulewidth{\Origarrayrulewidth}}%
}

\newcommand\Thickvrule[1]{%
  \multicolumn{1}{!{\vrule width 2pt}c!{\vrule width 2pt}}{#1}%
}

\begin{abstract}
Analysis of data related to software development helps to increase quality, control and predictability of software development processes and products. 
However, collecting such data for is a complex task. A non-invasive collection of software metrics is one of the most promising approaches to solve the task. In this paper we present an approach which consists of four parts: collect the data, store all collected data, unify the stored data and analyze the data to provide insights to the user about software product or process. We employ the approach to the development of an architecture for non-invasive software measurement system and explain its advantages and limitations. 

\end{abstract}

\section{Introduction}
Analysis of data related to software development helps to increase quality, control and predictability of both a development process and a resulting software product \cite{Vera-Baquero2013BusinessApproach}. Collecting such data gives an opportunity to reconstruct software development process and produce insights on how to improve it. However, collecting the data is a complex task \cite{ScottoSSV04}. An option is always collect the data ex-post trough questionnaires and qualitative (some times with a level of subjectivity) \cite{Tumyrkin}. However, a non-invasive collection of software metrics is one of the most promising approaches to solve the task \cite{Scotto2004,Janes2006}. 

There are systems which are targeting the area, but new available technologies, frameworks, libraries and tools enable a novel architecture for non-invasive measurement and analysis of software. Existing systems for non-invasive data collection typically use two types of metrics: software product metrics and software process metrics \cite{Fenton1998}. The data about software products and software development processes could be collected from developers' machines, smartphones, smart things, product repositories, task and defect tracking tools. The variety of sources and possible tools for data collection as well as many possible scenarios for data analysis make an issue of architectural design for developers of non-invasive software measurement systems. 


The main goal of this preliminary study is to establish basic approach and principles of system architecture for non-invasive software measurement systems. The contribution of the work is focused on three aspects: (i) an approach that guides design decisions; (ii) a set of core elements for such systems and (iii) an analysis of architectural decisions.  

In section \ref{sec:arch} we present the system architecture for non-invasive software metrics collection. In section \ref{sec:discussion} we discuss the architectural decisions made; and in section \ref{sec:usecase} we demonstrate a use case of the system. Sections \ref{sec:relatedwork} and \ref{sec:conclu} are devoted to related works and conclusions.


\section{An architecture for non-invasive metrics collection}
\label{sec:arch}
In this section, we will present the approach to collect and analyse data as well as the system architecture and the underlying technologies in use. Although systems for non-invasive data collection have been presened before (see section \ref{sec:relatedwork} for a comprehensive account), the approach presented in this paper is peculiar of this specific work, and represents one the major contribution of the study.

\subsubsection{Collect-Store-Unify-Analyze (CSUA) approach.}
This approach consists of four parts: first of all we \textit{collect} the data, such as metrics and events, from numerous distributed heterogeneous data sources. Second, we \textit{store} all collected data in raw format suitable for future use. The third part consists of data \textit{unifiers}, which can extract different relational data representations from non-relational stored data. Finally, we \textit{analyze} the data providing insights to the user about observed product or process.  

The CSUA approach guides the design of the architecture of a system for non-invasive software metrics collection.
The architecture developed according to the CSUA approach presented in Fig. 1. 
Basic purposes for such architectural design are collection, storage and analysis of metrics as well as flexible representation of data in dashboards. The core elements of the architecture are:
\begin{itemize}
\item  \textit{agents} for collecting data;
\item  \textit{databases}:  document-oriented and relational;
\item  data \textit{unifiers} and data \textit{exporters};
\item  \textit{dashboards} and applications for data analysis.
\end{itemize}

The following subsections describe these components and major data flows in the system.  

\begin{figure}
\centering
\label{fig:arch}
\includegraphics[width=90mm]{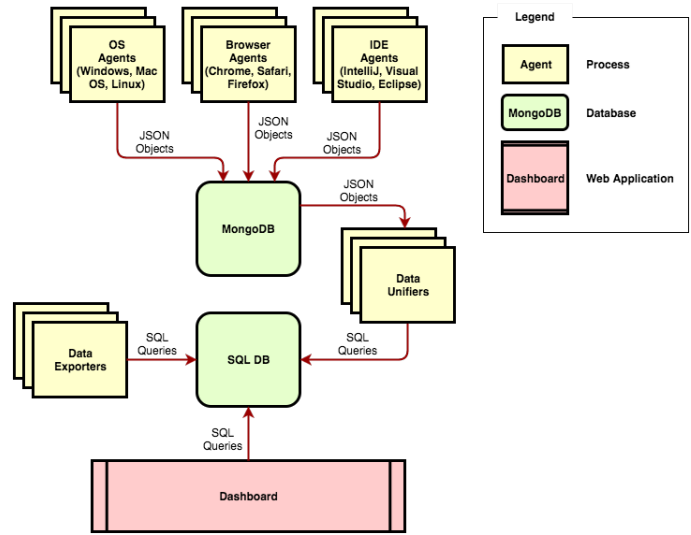}
\caption{Data flow in the system}
\end{figure}

\textbf{Agents.} A system for non-invasive software metrics collection gathers data about software products and software development processes. The data sources usually include developers' machines, smartphones and other devices; product repositories, task and defect tracking tools used in collaborative development. Data collection can be performed by multiple software agents of various types and kinds. The main purpose of an agent is data collection about a product and/or a process. There are multiple levels for of agents to operate and collect data (e.g. OS-level agents, browser-level agents, IDE-level agents, etc. 
 
OS agents are background operating system services for Windows, Linux, and Mac OS. Browser agents are extensions for Chrome, Firefox, and Safari. IDE agents are collecting data from Visual Studio, IntelliJ IDEA, Eclipse, XCode.
The system is not limited only to these types, we are planning to add agents for bug tracking systems, version control systems, etc. Agents are in an early development phase at the moment\footnote{User interface for one of prototype agents is shown in Fig. 2.}.
Moreover, data could be normalized in many different ways, but we do not want to force one common data schema to every agent, we will make this decision later (according to Lean Development principles) in the "Data Unifier" component.

\textbf{Document-oriented database.} To store data collected by agents we use document-oriented database -- MongoDB. The reasons why we chose this database is that it provides easy sharding of data, horizontal scaling and it uses JSON documents to store data. 

A connector between an agent and a document-oriented database works in the following manner.
An agent pushes data to a common document-oriented database over HTTP channel using RESTful API and JSON documents as data representation. We only impose a common high-level structure of JSON documents. A listing with example of a JSON document is provided below.

\scriptsize

\begin{lstlisting}[language=json,firstnumber=1]
{
  "timestamp": "2016-11-15T13:25:43.511Z",
  "agent": {
    "code_name": "MacOS developer's agent",
    "full_name": "Developer's activity collector",
    "secret_key": "6a81d622-5e24-4d9e-adc0-e3f7f2d93ac7",
    "install_guid": "2187b011-6b9d-4d86-8083-dd09a0d73019"
  },
  "metrics": {
    "event_id": "4a8acf6e7fbadc242de5b4f3",
    "event_type": "web-browsing",
    "event_duration": 1800,
    "user": {
          "username": "student",
          "company": "Innopolis University"
        },
    "host": {
          "host_name": "lab5_pc1",
          "ip_address": "10.90.121.49",
          "mac_address": "FF-FF-FF-FF-FF-FF",
          "os_version": "macOS 10 Sierra Version 10.12.1",
          "sw_version": "Safari Version 10.0.2 (12602.3.12.0.1)",
        },
     "sample_metric_data" : [
     	"stackoverflow.com", "google.com", "youtube.com"
     ]
  }
}
\end{lstlisting}


\normalsize

The document consists of three parts: 
\begin{enumerate}
\item Timestamp
\item Agent information
\item Collected metrics
\end{enumerate}

In the example, the top-level fields ``timestamp'' and ``agent''  describe the metadata, while the ``metrics'' part stores the actual data. The schema of the collected data may depend on an agent, but metadata fields stay the same across different agents. A sample user interface of an agent collecting process data is represented in Fig. 3. (section \ref{sec:usecase}).

\textbf{Data unifiers.} Data unifiers are processes which transform a set of JSON documents into rows and tables of a relational database. Resulting schema in each data unifier could be different depending on type of analysis that a customer may want to perform. Data unifiers pull data from MongoDB over HTTP channel using the same RESTful API as agents do.

\textbf{Relational database.} There could be multiple relational databases which our system may need to connect to. Hence, each data unifier serves as an adapter that write data to its own database. 

\textbf{Data exporters.} 
The architecture provides data exporter component, so users of the system could do their own analysis. Basically, data exporters convert data to several well-recognized formats, like csv-file, arff-file, etc.

\textbf{Dashboarding applications.} Dashboard is an application which supports decision making by simplifying the data and representing it a visual form. Backend part of a dashboarding application connects to a relational database. Frontend is rich with graphs, charts, and data visualization. 
A developer of dashboarding applications may have more details later, so our system should be ready to adapt to these changes. That's why modifiability of the system is highly demanded feature. 

\section{Discussion of architectural decisions}
\label{sec:discussion} 
In this section we discuss significant architectural decisions, what options we considered and why we chose the structures that have been presented above. These architectural decisions affect attributes of the system, therefore we discuss them together in Table 1.

Attributes such as extensibility, modifiability and consistency would benefit of a migration into the microservice paradigm \cite{ms-pause}. Recent projects of our team demonstrated an effective use and deploy of the paradigm in the field of ambient intelligence and smart buildings \cite{Salikhov2016a,Salikhov2016b}, in particular when associated with programming languages specifically designed with this purpose \cite{Bandura2016}, and with adequate programming abstractions \cite{Safina2016}. 

\begin{center}
\begin{table}
\caption{Architectural decisions and motivation behind them}
\label{T:attr}
\begin{tabular}{m{3cm}|m{9cm}}
\hline
\textbf{Attribute name} & \textbf{Arguments}  \\ \hline
Extensibility & Proposed architecture allows to add new agents and new analysis tools without downtime or reconfiguration. 
 \\ \hline
Security and Privacy & The system could be deployed in multiple organizations. 
So we need to provide reasonable authorization, roles and access restriction settings.  \\ \hline
Performance & We need high-performance on write. There could be thousands of agents trying to write their data into document-oriented database at the same time. \\ \hline
Consistency & We do not require strong consistency, eventual consistency should be fine.
 \\ \hline
Modifiability & We require high modifiability of database schema. \\ \hline
Scalability & We need horizontal scalability in terms of volume of data. \\ \hline
\end{tabular}
\end{table}
\end{center}

\section{Use case: a MacOS agent prototype}
\label{sec:usecase}
In this section, we show a common use case of the CSUA approach. We demonstrate such approach by the MacOS collector prototype. At the moment, only a prototype client-side application has been developed; it collects and transfers data for storage into the server (Fig. 2 and Fig. 3). 

\textbf{Step 1: Collecting data.} A MacOS agent collects data in background and can be stopped at any time (see Fig. 2). At any time, the user may review the collected data, apply a filter to collected records and submit them. This possibility to manually stop, review and filter data before a submission makes the application friendly to users (especially to those users, who may consider it a spyware).
\begin{figure}
\label{ui-1}
\centering
\includegraphics[width=45mm]{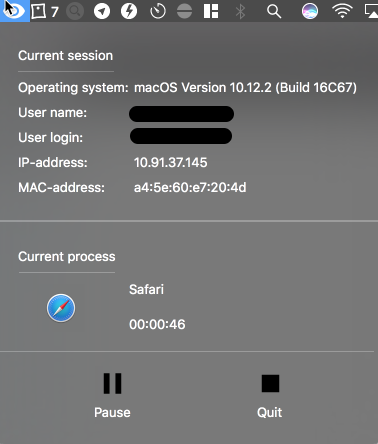}
\caption{User interface of a MacOS agent collecting data about user activity.}
\end{figure}
\begin{figure}
\centering
\includegraphics[width=95mm]{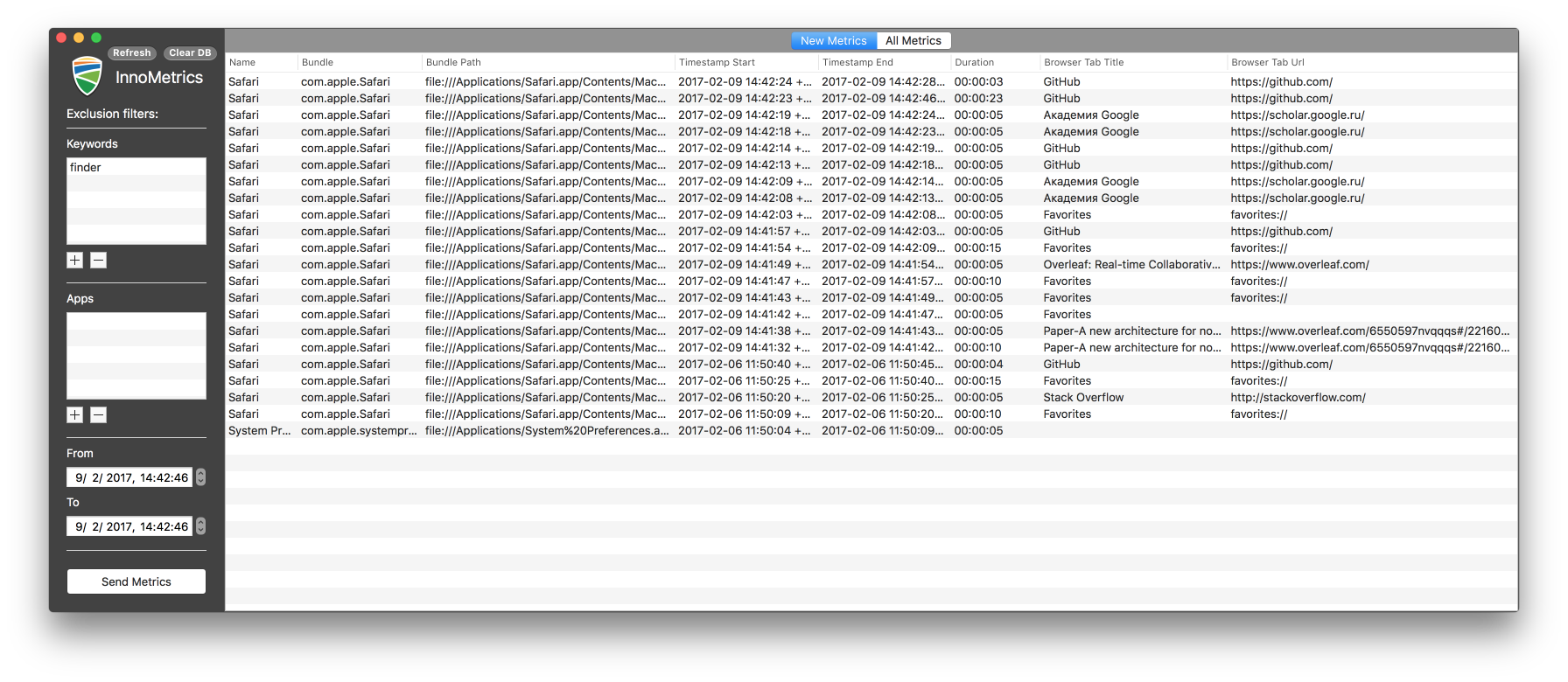}
\caption{User interface of a MacOS agent that represents collected data and transfers data to the server.}
\end{figure}

\textbf{Step 2: Store, filter and transfer data.} The interface for data transfer has several useful functions for accessing a collected dataset. A user may switch between newly collected (and not yet submitted) records and historical (submitted) records. In addition, there are three types of filters: a keyword filter, a filter by application and date/time filter (Fig. 3). 


\section{Related works in architectures for non-invasive measurement systems }
\label{sec:relatedwork} 
Over the past ten years several non-invasive measurement systems have been developed. In this section we review the following systems with emphasis on architecture: 
\begin{itemize}
\item PRO Metrics (PROM);
\item ElectroCodeoGram (ECG);
\item Empirical Project Monitor (EPM); 
\item Hackystat. 
\end{itemize}

\subsection{PRO Metrics }
PRO Metrics \cite{sillitti2003collecting,scotto2006non,sillitti2006managing} is a distributed architecture for collecting software metrics and Personal Software Process (PSP) data. PROM is based on Service-Oriented Programming development technique \cite{sillitti2002}. A client application stores collected information in XML file and does not deal with data transfer. This decision makes client-side components simpler. A transfer tool is separate client-application that transfers collected data and provides user authentication. 
Server-side components need to be installed and maintained only on one machine, therefore the overall complexity of the system is low. But in case of installation client components on many machines with different environment it becomes not a trivial task for a system administrator.

\subsection{ElectroCodeoGram} 
ElectroCodeoGram is a modular framework \cite{schlesinger2006electrocodeogram} aimed at micro-process research and discovering patterns in the sequence of events which describe the same programming behavior. For instance (i) copy and paste some piece of code with desired functionality and (ii) refactor code and make a function with needed parameters, represent two different patterns (or episodes) solving the same task. ECG supports micro-process research. It automatically records micro-process data using ECG Sensors; sends data to the central collection and analysis system. Data is transported over network sockets or SOAP. 

\subsection{Empirical Project Monitor}

Empirical Project Monitor \cite{ohira2004empirical,ohira2004empiricalsyst} is a system that automatically collects data (by ``pulling'') from three different repositories: 
\begin{itemize}
\item Configuration management systems;
\item Mailing list managers (e.g. Mailman, Majordomo);
\item Issue tracking systems (e.g. Bugzilla).
\end{itemize}
The EPM system consist of three components:
\begin{itemize}
\item Automatic data collection. EMP automatically collects data from repositories. 
\item Format translation and data store. EMP converts collected data to XML format. Converted data is stored in the PostgreSQL database. 
\item Analysis and visualization. EPM gets data for analysis from the database for visualization. 
\end{itemize}

\subsection{Hackystat}
Hackystat \cite{johnson2003beyond,johnson2007requirement,johnson2004practical} is a system for automatic collecting development metrics from sensors (attached to development tools). Hackystat sends data to the server where this data is analyzed. Its sensors are able to collect: 
\begin{itemize}
\item activity data (e.g. which file is under modification of developer); 
\item size data (e.g. lines of code);
\item defect data (e.g. number of pass/fail status of unit tests).
\end{itemize}
A developer should install one or more sensors to begin using Hackystat and then register with its server. In later versions of Hackystat its architecture has been criticized for growing complexity; developers made a decision to review the architecture and reimplement Hackystat in a service-oriented architecture (SOA). The main challenges for this revision were almost complete reimplementation of the system and the need for system developers to move to new architectural concept and libraries. 

\section{Conclusion and future work}
\label{sec:conclu} 
Non-invasive collection of software metrics demonstrated to be effective in the field of software measurement. Several systems for non-invasive data collection have been presented in the past. However, the approach presented in this paper is innovative for its own nature: for the peculiarity of the data flow and for the specific architecture adopted, as well as for the underlying architectural decisions. 
The architecture is designed to provide an high level of Scalability and Modifiability, as well as a direct way to extend the system with new types of agents.
Forthcoming steps include development of agents for operating systems (Windows and Linux), specific IDEs, popular browsers, version tracking systems, task tracking systems, and defect tracking systems.

Recent trends and development in the field of software architecture has shown an increasing attention towards the microservice architecture, 
which promises to help managing scalability, elasticity and robustness \cite{ms-pause}. It is under consideration the possibility to migrate from the current design to this new approach.  At the moment, there is no concrete work in this direction in the field of non-invasive collection, therefore it would represent an innovative trait of the system.

\bibliographystyle{plain}
\bibliography{bibliography}

\end{document}